\hoffset = -10mm
\baselineskip=6mm
\vglue1cm \line {\vbox{\halign{\hfill#\hfill\cr Nice INLN 03/08 \cr}}
\hfill\vbox{\halign{\hfill#\hfill\cr February 2003\cr}}} \vglue2cm
\magnification=1150

\centerline { {\bf ANGULAR INTRICACIES}}\bigskip\centerline{{\bf IN}}\bigskip
\centerline{{\bf HOT GAUGE FIELD THEORIES
}} 
 \bigskip\bigskip\bigskip \centerline {{${\rm{T.\ Grandou}} $ }}
\bigskip\medskip 
\centerline { {\it Institut Non Lin\'eaire de Nice UMR CNRS
6618; 1361, Route des Lucioles, 06560 Valbonne, France}}\centerline
{e-mail:Thierry.Grandou@inln.cnrs.fr}
\bigskip\bigskip\medskip\bigskip\bigskip \centerline{\bf ABSTRACT}
\bigskip\smallskip\medskip It is argued that in hot gauge field theories, "Hard Thermal Loops" leading order
calculations call for a definite sequence of angular averages and discontinuity (or Imaginary part prescription)
operations, and
run otherwise into incorrect results. The ten years old collinear singularity problem of hot $QCD$, provides a striking
illustration of that fate.
 \bigskip\bigskip\noindent PACS: 12.38.Cy, 11.10.Wx \bigskip\noindent
{{\bf Keywords:}} Hot $QCD$, Resummation Program, infrared, mass/collinear singularities,..

\vfill  \eject

{\bf{1. Introduction}}
\bigskip

The intrinsic non perturbative nature of non zero temperature  Quantum field Theories has long been recognized
[1].
Naive thermal perturbation theory can nevertheless be devised, both in imaginary and real time formalisms [2],
but
then, it promptly appears that, under certain circumstances, the original perturbative series must be
re-organized.
Such an example of re-organization is provided by the so called {\it{Resummation Program}} [3]. This program,
$RP$ for short, is a
resummation scheme of the leading order thermal fluctuations which, in the literature, are known under the spell
of
{\it{Hard Thermal Loops}}. Whenever one is calculating a physical process related to thermal Green's functions
whose external/internal legs are {\it{soft}}, it is mandatory to trade the naive thermal
perturbation theory for the $RP$. The softness alluded to above, refers to momenta on the
order of the soft scale $gT$, where $T$, the temperature, stands for the {\it{hard}} scale and $g$ for any relevant
(bare/renormalized) and small
enough coupling constant.\medskip
The $RP$ which has been set up in order to remedy an obvious lack of completeness of the naive thermal
perturbation theory, has produced interesting results, but has also met serious obstructions in the infrared
regime
of the theories [4],[5]. Within the Resummation Program itself, the solutions proposed so far [5,6,7], however
interesting in their own respect, cannot be organized a systematic way and display a crucial dependence on the
process under consideration.\smallskip In the past few years, another resummation scheme of the leading order
thermal fluctuations has been proposed
by the present author, and seems to avoid all of the infrared problems met by the standard $RP$, [8]. In loop
calculations (where both resummation schemes are
being used as effective perturbation theories, ruling the leading thermal fluctuations) the former only differs
the
standard $RP$, the sequence along which are performed an integration over a looping energy $p_0$, on the one
hand,
and a sum over the number $N$ of $HTL$-self energy insertions, on the other hand. This new resummation scheme,
has accordingly been denoted $P\!R$ for short, and is, by
construction, a Perturbative Resummation scheme of the leading order thermal fluctuations. \smallskip Such is not the
case of the $RP$. While $HTL$ $n$-point
vertices are purely perturbative objects, effective propagators are not, giving rise to pole residues and dispersion
laws that
cannot be obtained out of pure thermal perturbation theory. In the infrared limit, in particular, those effective
propagators are known to give rise to series in the coupling constant that are no longer
Taylor's but Laurent's series [8]. In any higher number of loops calculations, this simple, still crucial
difference,
allows one to foresee easily, why, in the $RP$ case
and not in the $P\!R$ one, some {\it {enhancement mechanisms}} may occur (and effectively will!) so as to
make higher number of loops as important, and even bigger, than lower number of loops diagrams [9].
\bigskip In this note, we take advantage of a comparison of the two $RP$ and $P\!R$ resummation
schemes to point out some overlooked aspects concerning "the historical derivations" of the famous {\it{collinear
singularity problem of hot $QCD$}}, [5]. As many of our further calculations have displayed ever since, these
aspects are generic and extend beyond the historical derivations to be recalled shortly.\smallskip
We
will be using the convention of upper case letters for quadrimomenta and lower case ones for their components,
writing, for example $P\!=\!(p_0, {\vec p})$. Our conventions for labelling internal and external momenta can be
read
off Fig.1. \bigskip
{\bf{2. The collinear singularity problem of hot $QCD$}}\medskip
This almost ten years old issue is the following. The soft real photon emission rate out of a Quark-Gluon Plasma in
thermal equilibrium
involves, in particular, the calculation of the quantity 
$$\displaylines{\Pi_R(Q)=i\int {{\rm d}^4P\over
(2\pi)^4}(1-2n_F(p_0))\ {\rm
disc_P}\ Tr\biggl\lbrace {}^\star S_R(P)\  {}^\star\Gamma_\mu(P_R,Q_R,-P'_A)\cr\hfill {}^\star S_R
(P')\ {}^\star\Gamma^\mu(P_R,Q_R,-P'_A)\biggr\rbrace\qquad(2.1)}$$The discontinuity is to be taken in the
energy variable $p_0$, by forming the
difference of $R$ and $A$-indiced $P$-dependent quantities, and within standard notations, fermionic
{\it{HTL}}
self energies, effective propagators and vertices are respectively given by

$${}^\star S_\alpha(P)={i\over {\rlap / \!P} - \Sigma_\alpha(P)+i\epsilon_\alpha p_0}\ ,\ \ \ \ \alpha=R,A\ ,\ \ \ \
\epsilon_R=-\epsilon_A=\epsilon
 \eqno(2.2)$$

$$\Sigma_\alpha(P)=m^2\int {{\rm{d}}{\widehat K}\over 4\pi} {{{\rlap / \!\widehat K}} \over {{ \widehat
K}}\!\cdot\! P+i\epsilon_\alpha}\ ,\ \ \ \ m^2= C_F{g^2T^2\over 8}\eqno(2.3)$$

$${}^\star\Gamma_\mu(P_\alpha,Q_\beta,P'_\delta)=-ie\left(\gamma_\mu+\Gamma^{HTL}_\mu(P_\alpha,Q_\beta,P'_\delta)\right)\eqno(2.4)$$

$$\Gamma^{HTL}_\mu(P_\alpha,Q_\beta,P'_\delta)=m^2\int{{\rm d}{\widehat
K}\over 4\pi}{ {\widehat k}_\mu\  {{\rlap / \!\widehat K}}\over
({\widehat K}\!\cdot\! P+i\epsilon_\alpha)({\widehat
K}\!\cdot\! P'+i\epsilon_\delta)}\eqno(2.5)$$where ${\widehat
K}$ is the lightlike four vector $(1,{\widehat
k})$. In view of (2.4), four terms come about, three of them proportional to a collinear singularity. These
singular terms are the two terms with one bare vertex $\gamma_\mu$, the other $\Gamma^{HTL}_\mu$, plus the
term including two $HTL$ vertices, $\Gamma^{HTL}_\mu$. Thanks to an abelian Ward identity peculiar to the high
temperature limit, a partial cancellation of these collinear singularities occurs, but out of the term including two
$\Gamma^{HTL}_\mu$ vertices, a
collinear
singularity remains,
$$\displaylines {-{2i{e^2m^2\over q^2}}\left( \int{{\rm d}{\widehat
K}\over 4\pi}{1\over {\widehat
Q}\!\cdot\! {\widehat
K}+i\epsilon}\right)
\int {{\rm d}^4P\over
(2\pi)^3}\ \delta(P\!\cdot\!{\widehat Q})\ (1-2n_F(p_0))
\cr\hfill \times\  
[Tr\left({}^\star S_A(P){{\rlap / \!\widehat Q}}\right)-Tr\left({}^\star S_R(P'){{\rlap / \!\widehat
Q}}\right)]\qquad(2.6) }$$where, the soft photon being real, $Q$ is the lightlike $4$-vector
$Q\! =q {\widehat{Q}}=\! q (1,{\widehat{q}})$. In the literature, this result is written in the form
$$ {C^{st}\over\varepsilon}\int
{{d^4P}\over (2\pi)^4}\ \delta({\widehat{Q}}\!\cdot\! P)\ (1-2n_F(p_0))\sum_{s=\pm
1,V=P,P'}\pi(1-s{v_0\over v})\beta_s(V) \eqno(2.7)$$where the overall $1/\varepsilon$ comes from a dimensionally
regularized evaluation of the factored out angular integration appearing in (2.6), and where $\beta_s(V)$ is related
to
the effective fermionic propagator usual parametrization [2],
$${}^\star S_{\alpha}(P)={i\over 2}\sum_{s=\pm 1}{{{\rlap / \!\widehat
P_s}}}\Delta_s(p_0+i\epsilon_\alpha, p)\eqno(2.8)$$with ${\widehat{P_s}}=(1,s{\widehat{p}})$, the
label
$s$ referring to the two dressed fermion propagating  modes. One has then
$$\Delta_s(p_0+i\epsilon_\alpha, p)\equiv \Delta^\alpha_s(p_0, p)=\alpha_s(p_0,p)- i\pi\epsilon(\epsilon_\alpha)
\beta_s(p_0,p)\eqno(2.9)$$where $\epsilon(x)$ is the distribution "sign of $x$".\bigskip\bigskip\bigskip {\bf{3. On
an
improper derivation }}\bigskip The third term, with the two $HTL$ vertices, $\Gamma^{HTL}_\mu$, involves two angular
integrals $W_i$ that are,
with $i\in \{1,2\}$,
$$W_i(P,P')=\int{{\rm d}{\widehat K}\over 4\pi}\int{{\rm d}{\widehat
K'}\over 4\pi}\ {({\widehat
K}\!\cdot\!{\widehat
K'})^i\over ({\widehat K}\!\cdot\! P+i\epsilon)({\widehat
K}\!\cdot\! P'+i\epsilon)({\widehat K'}\!\cdot\! P+i\epsilon)({\widehat
K'}\!\cdot\! P'+i\epsilon)}\eqno(3.1)$$While $W_2$ only is met in the standard $RP$ calculation, $W_1$ and $W_2$ come
into play within the $P\!R$ scheme.
As shown in [10], these integrals are given by very cumbersome expressions, so that, following ~[5], we will
illustrate
our point on the two diagrams including one bare vertex  $\gamma_\mu$, the other $\Gamma^{HTL}_\mu$. The $R/A$ real
time formalism conveys us to the expression
$$\displaylines{ \Pi^{(\star,\star;1)}_R(Q)= -ie^2m^2\int {{\rm d}^4P\over
(2\pi)^4}(1-2n_F(p_0))\cr\hfill {\rm
disc_P}\int{{\rm d}{\widehat
K}\over 4\pi}   {\ Tr\left({}^\star S_R(P){{\rlap / \!\widehat K}} {}^\star S_R(P'){{\rlap / \!\widehat
K}}\right)\over
({\widehat K}\!\cdot\! P+i\epsilon)({\widehat
K}\!\cdot\! P'+i\epsilon)}\qquad(3.2)} $$where the superscript $(\star,\star;1)$ in the left hand side refers to a
self
energy diagram involving two effective propagators and one vertex $HTL$ correction. The steps leading to the
collinear singularity of (3.2) are as follows (the second reference of [5], using the $R/A$ real time formalism, is
followed here). The discontinuity in $p_0$ is taken in consistency with the $R/A$ formalism, by writing

$$\displaylines{{\rm
disc_P}{\ Tr\left({}^\star S_R(P){{\rlap / \!\widehat K}} {}^\star S_R(P'){{\rlap / \!\widehat K}}\right)\over
({\widehat K}\!\cdot\! P+i\epsilon)({\widehat
K}\!\cdot\! P'+i\epsilon)}={\ Tr\left({}^\star S_R(P){{\rlap / \!\widehat K}} {}^\star S_R(P'){{\rlap /
\!\widehat K}}\right)\over
({\widehat K}\!\cdot\! P+i\epsilon)({\widehat
K}\!\cdot\! P'+i\epsilon)}\cr\hfill -{\ Tr\left({}^\star S_A(P){{\rlap / \!\widehat K}} {}^\star S_R(P'){{\rlap /
\!\widehat K}}\right)\over ({\widehat K}\!\cdot\! P-i\epsilon)({\widehat
K}\!\cdot\! P'+i\epsilon)}\qquad(3.3)}$$Closing the $p_0$ integration contour in the upper half complex $p_0$
plane,
one selects a pole term coming from the vertex $\Gamma^{HTL}_\mu$ which reads
$$\displaylines{ \Pi^{(\star,\star;1)}_R(Q)= -e^2m^2\int {{\rm d}^4P\over
(2\pi)^3}(1-2n_F(p_0))\cr\hfill \int{{\rm d}{\widehat
K}\over 4\pi}\ {\delta({\widehat K}\!\cdot\! P)\over {\widehat
K}\!\cdot\! P'+i\epsilon}   \ Tr\left({}^\star S_A(P){{\rlap / \!\widehat K}} {}^\star
S_R(P'){{\rlap / \!\widehat K}}\right)
\qquad(3.4)} $$Since $P'\!=\! P+Q$, we have indeed, as a building bolck of (3.4), the expression
$$\int{{\rm d}{\widehat
K}\over 4\pi}\ {\delta({\widehat K}\!\cdot\! P)\over {\widehat
K}\!\cdot\! Q+i\epsilon}   \ Tr\left({}^\star S_A(P){{\rlap / \!\widehat K}} {}^\star
S_R(P'){{\rlap / \!\widehat K}}\right)\eqno(3.5)$$The angular average therefore develops a collinear singularity in a
neighbourhood of
${\widehat
K}\!\!=\!\!{\widehat Q}$, not compensated for
by
the numerator, and whose "residue" reads
$${C^{(st)}\over \varepsilon}\ {\delta({\widehat Q}\!\cdot\! P)\over q}Tr\left({}^\star S_A(P){{\rlap /
\!\widehat
Q}} {}^\star S_R(P'){{\rlap / \!\widehat Q}}\right)\eqno(3.6)$$The first factor, singular at $\varepsilon
=0$,
is obtained by using a dimensional regularization of the angular integral. Up to regular
contributions that we do not consider here,
this singular piece of (3.5) translates, for $\Pi^{(\star,\star;1)}_R(Q)$, into the singular result
$$\displaylines{ \Pi^{(\star,\star;1)}_R(Q)= -{C^{(st)}\over \varepsilon}{e^2m^2\over q}\int {{\rm d}^4P\over
(2\pi)^3}(1-2n_F(p_0))\cr\hfill \times\ \delta({\widehat Q}\!\cdot\! P)\ Tr\left({}^\star S_A(P){{\rlap /
\!\widehat
Q}} {}^\star S_R(P'){{\rlap / \!\widehat Q}}\right)
\qquad(3.7)} $$The contribution of (3.7) to the soft photon emission rate being proportional to its imaginary part,
the emission rate is thus plagued with a collinear singularity. In its original (published) version [5], this result
is
obtained within the imaginary time formalism, continued to real energies, where the $RP$ has been
first devised [3].\medskip In a recent article [10], the analogous $P\!R$ calculation has been proven
to
be mass/collinear singularity free. In particular, such is the case of any of the contributions to the soft real
photon emission rate
due to $\Pi^{(N,N';1)}_R(Q)$ quantities, the sum of which, on $N$ and $N'$ corresponding to
$\Pi^{(\star,\star;1)}_R(Q)$, as depicted on Fig.1. Thus, at face value, a collinear singularity shows up in a
resummation scheme, not in the other. Let us now examine how can this be so.
\medskip

In a $P\!R$ scheme, one has indeed for the potentially dangerous most part of any $\Pi^{(N,N';1)}_R$ quantity,
the
expression [10], $$\displaylines{-ie^2m^2(\ ..\ )\sum_{n,n'}(m^2)^{(N'+N-2n'-2n)}\int {{\rm d}^4P\over
(2\pi)^4}(1-2n_F(p_0))\ {1\over p'p}\left({[-{P'}^2\Sigma^2_R(P')]^{n'}\over ({P'}^2+i\epsilon
p'_0)^{N'}}\right)\cr\hfill {\rm{disc_P}}\ \left({[-{P}^2\Sigma^2_R(P)]^{n}\over ({P}^2+i\epsilon
p_0)^{N}}\right)\ \ \int{{\rm d}{\widehat
K}\over 4\pi}\ {1\over {\widehat
K}\!\cdot\! P+i\epsilon}\ {1\over {\widehat K}\!\cdot\!
P'+i\epsilon}\qquad(3.8)}$$where the  discrete sums on $n$ and $n'$ extend over a finite set of integers and need not
be furhter specified here. Performing the angular integration, one finds 
$$\int{{\rm d}{\widehat
K}\over 4\pi}\ {1\over {\widehat
K}\!\cdot\! P+i\epsilon}\ {1\over {\widehat K}\!\cdot\!
P'+i\epsilon}={1\over 2Q\!\cdot\!P+i\epsilon q}\ln{P^2+2Q\!\cdot\!P+i\epsilon p'_0\over P^2+i\epsilon
p_0}\eqno(3.9)$$ 
The collinear domain is included in the phase space region where ${\widehat Q}\!\cdot\! P\simeq 0$, where one has

$${1\over 2Q\!\cdot\!P+i\epsilon q}\ln{P^2+2Q\!\cdot\!P+i\epsilon p'_0\over P^2+i\epsilon p_0}= {1\over
P^2+i\epsilon p_0}-{1\over 2}{2Q\!\cdot\!P+i\epsilon q\over
(P^2+i\epsilon p_0)^2}+..\eqno(3.10)$$Obviously, this $HTL$-vertex induced behaviour is potentially mass singular by
the light
cone region $P^2\!\simeq\! 0$. This means that in a neighbourhood of ${\widehat Q}\!\cdot\! P\simeq 0$, (3.9)
mixes up with partial propagators, $S_R^{(N(N'))}(P(P'))$, own
potentially mass singular behaviours. Recalling that we have
$$S^{(N)}_R(P)={i\rlap / \!P\left(\rlap /\!\Sigma_R(P)\rlap /\!
P\right)^{N}\over (P^2+i\epsilon p_0)^{N+1}}\eqno(3.11)$$the whole integrand structure of (3.8), in the collinear
domain,
is readily seen to "boil down" to a simple shift of inverse power $${1\over (P^2+i\epsilon p_0)^{M}}\  \longmapsto\
{1\over
(P^2+i\epsilon
p_0)^{M+1}}\ ,\ \ \ \
M=N+N'\eqno(3.12)$$As demonstrated in [10], similar shifts, as well as
many more transformations, are proven to leave totally unaffected the robust mass singularity cancellation
patterns which guarantee the regular character of the $P\!R$ calculation. \smallskip That is, in
contradistinction with
the $RP$ scheme, a picture emerges out of the $P\!R$ scheme, where effective vertices {\it{potentially}} mass
singular behaviours, melt with partial
effective propagators own {\it{potentially}} mass
singular behaviours
into structural patterns which rule the overall compensation of {\it{actual}} mass/collinear singularities, [8,10]. 
\bigskip
However, it is important to remark that, apart from their constitutive difference, not exactly the same
steps have been
followed in either $RP$ and $P\!R$ schemes. Both resummation schemes are here being developed within the $R/A$ real
time formalism, so that, starting from the $P\!R$ expression (3.8), exactly the same
procedure as followed from (3.3) to (3.7)
may be applied, with, as a straightforward result
$$\displaylines{-{C^{(st)}\over \varepsilon}{e^2m^2\over q}(\ ..\ )\sum_{n,n'}(m^2)^{(N'+N-2n'-2n)}\int {{\rm
d}^4P\over
(2\pi)^3}(1-2n_F(p_0))\ {1\over p'p}\cr\hfill \times\ 
\delta({\widehat Q}\!\cdot\! P)\  {[-{P'}^2\Sigma^2_R(P')]^{n'}\over ({P'}^2+i\epsilon
p'_0)^{N'}}{[-{P}^2\Sigma^2_A(P)]^{n}\over ({P}^2-i\epsilon
p_0)^{N}}\ +\ ..\qquad(3.13)}$$In words, (3.8) is now discovered to display the same collinear singularity as the
one plaguing the $RP$ result (3.7), in sharp contradiction with our claim that any
$\Pi^{(N,N';1)}_R(Q)$ contributes
mass/collinear singularity free quantities to the soft photon emission rate.\smallskip
The contradiction is of course only apparent and entirely rooted in the fact that passing from (3.1) to (3.3), the
prescription of discontinuity in $p_0$ has been  commuted with the angular integration on ${\widehat K}$. As we have
just come to see, in both $RP$ and $P\!R$
schemes, the effect of that improper commutation (the discontinuity of the integral is prescribed by the
formalisms,
and not the integral of the discontinuity!)
is to provide the collinear singularity with a somewhat {\it{absolute}} status. In either schemes in effect, the
collinear singularity just factors out, in total independence of the remaining
integrations to be performed.

The lack of commutativity of the discontinuity and angular integration can be the matter of a direct observation.
From
(3.9), one has $$\displaylines{{\rm{disc_P}}\int{{\rm d}{\widehat K}\over 4\pi}\ {1\over {\widehat
K}\!\cdot\! P+i\epsilon}\ {1\over {\widehat K}\!\cdot\!
P'+i\epsilon}={i\pi\epsilon(p_0)\Theta(-P^2)\over { Q}\!\cdot\! P+i\epsilon q}-{i\pi\over q}\delta({\widehat
Q}\!\cdot\! P)\ln{P^2+2Q\!\cdot\!P+i\epsilon p'_0\over
P^2+i\epsilon
p_0}\cr\hfill  = {i\pi\epsilon(p_0)\Theta(-P^2)\over { Q}\!\cdot\! P+i\epsilon  }\qquad(3.14)}$$In the right hand
side, the term proportional to the distribution $\delta({\widehat Q}\!\cdot\! P)$, has coefficient zero, and not
the collinear singularity ${C^{(st)}/ \varepsilon}$, as historically derived whithin the
reverse,
improper sequence. Though the illustration above, (3.5)-(3.6), is given within the $R/A$ real time formalism only,
it must
be noticed that, {\it {mutatis mutandis}},
both the historical derivations of [5] hinge on a reverse sequence calculation,  $$\int{{\rm d}{\widehat K}\over
4\pi}\ \left(\{{\rm{disc_P}}\ ,\ {\rm{or}}\  {\cal{I}}m \}\ {1\over {\widehat K}\!\cdot\!
P+i\epsilon}\right)
\ {1\over {\widehat K}\!\cdot\!
P'+i\epsilon}=-{i}\ \{{1\over 2}\ , {\rm{or}}\  {1\over 4}\}\int{{\rm d}{\widehat
K}}\ {\delta({\widehat K}\!\cdot\! P)\over {\widehat K}\!\cdot\! Q +i\epsilon}\eqno(3.15)$$along which a collinear
singularity
is identified at ${\widehat K}\!\!=\!\!{\widehat Q}$. Then, as advertised after (2.7), one relies on a dimensionally
regularized angular integration to find out the singular counterpart 
$$-{i}\ \{{1\over 2}\ , {\rm{or}}\  {1\over 4}\}\int{{\rm d}{\widehat
K}}\ {\delta({\widehat K}\!\cdot\! P)\over {\widehat K}\!\cdot\! Q +i\epsilon}=-{i}{C^{(st)}\over
\varepsilon}\delta({\widehat Q}\!\cdot\! P)\eqno(3.16)$$ where $\varepsilon$ is the regularizing parameter of an
angular integration performed at
$D\!=\! 3+2\varepsilon$ spatial dimensions. This derivation, however, is inaccurate. This can be verified by a direct
calculation of (3.15)'s right hand side, which gives instead

$$\int{{\rm d}{\widehat K}\over 4\pi}\ \left({\rm{disc_P}}\ {1\over {\widehat
K}\!\cdot\! P+i\epsilon}\right) {1\over {\widehat K}\!\cdot\!
P'+i\epsilon} ={i\pi \Theta(-P^2) \delta({\widehat{p}}-{\widehat{q}})\over { Q}\!\cdot\!
P-i\epsilon
}\eqno(3.17)$$ A first important remark is
that under subsequent angular (${\widehat{p}}$) and energy ($p_0$)
integrations, the above reverse sequence result (3.17) will lead to more singular expressions than the proper
sequence one, (3.14). Likewise, the correct prescription of $+i\epsilon$ is not preserved by the sequence
(3.17). Now, the point is that nothing like the "absolute", overall factoring out collinear singularity of (3.16),
i.e. the
term ${C^{(st)}/ \varepsilon}$, ever appears in the final
result, and this underlines
the illicit character of the manipulations leading to it.\smallskip Discarding thus (3.16), one may still compare
the right hand sides of (3.14) and (3.17). Then, apart from the fact that a non vanishing result is obtained for
strictly collinear ${\vec p}$ and ${\vec q}$ momenta only, 
another essential difference comes about immediately, which is the sign
distribution $\epsilon (p_0)$. In real time formalisms, sign distributions have long been noticed to be
crucial so as to preserve integrable and non-integrable mass singularity compensations [11], and as observed
here again, they are a natural outcome
of the calculational operations proper sequence (3.14).\smallskip  Indeed, this is the very place where
intricacies are taking place, between angular averages peculiar to $n$-points $HTL$-vertices, and discontinuity or
Imaginary part
prescriptions. Staring at (3.9), for example, one can observe how the angular average is able to
correctly reproduce all of
the relevant internal
legs $R/A$ specifications, the $i\epsilon q$, $i\epsilon p_0$ and $i\epsilon p'_0$, which do not appear at all in
the
left hand side. As can be read off [10],
sections 4 and 5, this is not an isolated situation and rather reveals to be general a mechanism.\smallskip That is,
the proper sequence of angular average and discontinuity operations complies
with the required, exact sign distributions which otherwise, are clearly endangered.\smallskip By the way, this
elucidates also the intriguing point made in the third reference of [8], after equation (3.32), which was then left
as
an issue. There, the discontinuity in $p_0$ of the angular integral $W_2$ was taken to be
$${\rm
disc}\ W_2(P,P')=-4i\pi\epsilon(p_0)\int{{\rm d}{\widehat
K}\over 4\pi}\ {\delta({\widehat K}\!\cdot\! P)\over {\widehat K}\!\cdot\! P'+i\epsilon} \int{{\rm d}{\widehat
K'}\over 4\pi}\ {({\widehat K}\!\cdot\!{\widehat K'})^2 \over 
({\widehat K'}\!\cdot\! P+i\epsilon)({\widehat
K'}\!\cdot\!
P'+i\epsilon)}$$whereas it is manifest that in conformity with the common use, the reverse improper sequence had been
followed. In order to preserve mass singularity cancellations though, the correct sign distribution, $\epsilon(p_0)$,
had to be restored by hand (this restoration was motivated a heuristic way, by recalling that the vertex corrections
$\Gamma_\mu^{HTL}$, were after all nothing but leading order approximations of full order $g^2$ expressions, endowed
with explicit $i\epsilon p_0$ specifications).\smallskip

 Since further, higher
number of loops calculations [12] enforce this peculiarity too, it seems reasonable to conclude that in any $RP$ or
$P\!R$ calculations, and any real or imaginary time formalism, angular
averages should definitely be performed first.

\bigskip\bigskip\bigskip
{\bf{4. The collinear problem and the proper sequence}}\bigskip
A full, extensive treatment of the soft real photon emission rate $RP$ calculation, as the proper sequence is
followed,
falls far beyond the scope of the short present note, and could be postponed to a future publication. We will here
content ourselves with an illustration
of how the collinear singularity problem gets translated when the angular integration is performed first.\smallskip
From (3.2) and (2.8), one obtains
$$\displaylines{ i{\cal{I}}m\ \Pi^{(\star,\star;1)}_R(Q)=-ie^2m^2\int {{\rm d}^4P\over
(2\pi)^2}(1-2n_F(p_0))\ {1\over pp'}\sum_{s,s'=\pm 1}ss'\beta_{s}(p_0,p)\beta_{s'}(p'_0,p')\cr
 +\ {e^2m^2\over 2}\int {{\rm d}^4P\over
(2\pi)^3}(1-2n_F(p_0))\ {1\over pp'}\cr\hfill \times\ \sum_{s,s'=\pm 1}\left({s'}\beta_{s'}(p'_0,p')\ (1-s{p_0\over
p}){\ \rm disc_P}\ {\Delta_R^s(P)}\ln({p_0+p\over p_0-p})\ +\ (P\leftrightarrow P')\right)\cr\hfill 
+\ e^2m^2\int {{\rm d}^4P\over
(2\pi)^3}(1-2n_F(p_0))\sum_{s,s'=\pm 1}\ (1-s'{p'_0\over p'})\ {\beta_{s'}(p'_0,p')}\cr\hfill \times\ (1-s{p_0\over
p}){\ \rm
disc_P}\ {\Delta_R^s(P)}\ {1\over 2Q\!\cdot\!P }\ \ln{P'^2\over P^2
}\qquad(4.1)} $$where the distribution $\beta_s(p_0,p)$ of (2.9), though textbook material, may be worth
recalling
here in view of the forthcoming comments   $$\displaylines{\beta_s(p_0,p)= Z_s (p)\delta\left(p_0-\omega_s
(p)\right)+Z_{-s} (p)\delta\left(p_0+\omega_{-s} (p)
\ \right)
\cr\hfill +
{ m^2\over 2p}{(1-s{p_0\over p})\Theta(-P^2)\over \left(p(1-s{p_0\over
p})-{m^2\over 2p}\left((1-s{p_0\over p})\ln|{p_0+p\over p_0-p}|+2s\right)\right)^2+{\pi^2 m^4\over
4p^2}(1-s{p_0\over p})^2}
\qquad(4.2)}$$\smallskip It is easy to check that the first line of (4.1), with
the two functions $\beta_s$, is safe. For example, one may rely on the integration contour technique used for energy
sum rules, [2], chapter 6, to show that $p_0$ and ${\widehat{p}}$ integrations are well defined. Likewise, inspection
shows that
no problem is inherited from the second and third lines of (4.1). At $p_0\!=\!\pm p$, in effect, the potentially
dangerous
term of $ln|{p_0+p/ p_0-p}|$ is, according to the value of $s=\pm 1$, either depleted by its own squared expression,
or
cancelled by a factor of
$(1-sp_0/p)^2$.\smallskip For the last term of (4.1), a potentially singular most contribution to ${\cal{I}}m\
\Pi^{(\star,\star;1)}_R(Q)$ is generated by taking the discontinuity in $p_0$ of the
first factor, $\Delta^s_R(P)$. One gets $$\displaylines{  -e^2m^2\int {{\rm d}^4P\over
(2\pi)^2}(1-2n_F(p_0))\sum_{s,s'=\pm 1}\ (1-s'{p'_0\over p'})\ {\beta_{s'}(p'_0,p')}\cr\hfill \times\ (1-s{p_0\over
p})\ {\beta_{s}(p_0,p)}\ {1\over 2Q\!\cdot\!P }\ \ln{P'^2\over P^2
}\qquad(4.3)} $$The Dirac pieces of both spectral densities $\beta_s$ and $\beta_{s'}$, pose no problem, whereas the
other parts, proportional to the Heaviside distributions $\Theta(-P^2)$ and $\Theta(-P'^2)$, and hereafter denoted
by ${\widehat{\beta}}_{s,s'}$, yield
$$\displaylines{ 
-e^2m^2\int {{\rm d}^3p\over
(2\pi)^2}\int_{-p}^{+p} {\rm d}p_0\ (1-2n_F(p_0))\ \sum_{s'=\pm 1}\ (1-s'{p'_0\over p'})\ {\widehat{
\beta}}_{s'}(p'_0,p')\cr\hfill \times  \sum_{s=\pm 1}(1-s{p_0\over p})\  {\widehat{\beta}}_{s}(p_0,p)\   {1\over
2Q\!\cdot\!P } \ln{P'^2\over P^2 }\qquad(4.4)}$$ 
In
the phase space domain where $2Q\!\cdot\!P$ is vanishing, the integrand of (4.4) behaves like
 $$ 
 (1-2n_F(pz))\sum_{s'=\pm 1}\ (1-s'{q+pz\over p'})\ {\widehat{
\beta}}_{s'}(q+pz,p')\times\ {-1\over p^2}\ \sum_{s=\pm 1}{\widehat{\beta_s}}(pz,p)\ {1\over 1+sz}\eqno(4.5) $$where
we
have introduced the scaling variable $z=p_0/p$, and where $p'^2(z)=p^2+2qpz+q^2$. Writing
$$1-2n_F(pz)=2[n_F(p)-n_F(pz)]+\left(1-2n_F(p)\right)\eqno(4.6)$$the last term only, in the right hand side, can
possibly induce a singularity by the light cone region ($z=\pm 1$), reached from below, and this would be singular
contribution can be expressed as
$$\displaylines{ 
{e^2m^2\over 2\pi}\Theta(-P^2)\int {p\ {\rm d}p}\ (1-2n_F(p))\int_0^1 {\rm d}z\  {{\widehat{\beta}}_{-}(pz,p)\over
1-z}\cr\hfill
\sum_{\eta=\pm 1}\left(1+{q+\eta pz\over p'(\eta z)}\right)\   \eta
{{\widehat{\beta}}_{-}\left(q+\eta pz,p'(\eta z)\right)}\qquad(4.7)} $$where use has been made of the relation
$\beta_s(-p_0,p)=\beta_{-
s}(p_0,p)$. Now, against all odds, the light cone ($1\!-\!z\simeq 0$) behaviours of ${\widehat{\beta}}_{-}(pz,p)$,
$\left(1+{q+\eta pz\over p'}\right) $, and
${{\widehat{\beta}}_{-}\left(q+\eta pz,p'(\eta z)\right)}$, make it straightforward to check that the above
potentially
dangerous most contribution is a regular one indeed.\smallskip
Since, out of the last term of (4.1), no other sign of a possible singularity shows up, it appears that the two
$\Pi^{(\star,\star;1)}_R(Q)$
contributions to the soft real photon emission rate are regular, in contradistinction with the usual, singular
results.

\eject
{\bf{5. Conclusion}}\bigskip
In this note, we have stressed that whatever the formalism (real or imaginary time) and the resummation
scheme (standard ($RP$) or perturbative ($P\!R$)), a definite sequence of angular averages and discontinuity (or
Imaginary part) operations, must definitely be preserved. Moreover, this proper sequence is nothing exotic, as it
simply corresponds to the sequence naturally prescribed by any real or imaginary time formalism one may use.
\smallskip 

Ignoring that proper sequence may lead (and has led) to incorrect derivations and/or results.
In
particular, infrared singularity cancellation patterns have long revealed to be particularly sensitive to the angular
averages and discontinuity operations proper sequence. This is also
a point
which, we think, should be kept in mind within the context of numerical simulations.\smallskip

For the hot $QCD$ collinear problem met in the $RP$ scheme, this point
certainly matters a lot. Our rapid analysis of the one effective vertex topologies, $\Pi^{(\star,\star;1)}$, has not
allowed us to detect any relic of a
mass/collinear singularity in the soft photon emission rate, the proper sequence being followed.\smallskip Now, as advertised
in section 3,
this indicates only that the proper sequence is less singular, under subsequent integrations, than the reverse
improper
one, and let totally open the issue concerning the two
effective vertex topology, $\Pi^{(\star,\star;2)}$, contribution to the emission rate.\smallskip
 In other words, the whole correct
settings of the hot $QCD$
collinear singularity $RP$-problem.. if any! .. has to be worked out again.

\vfill\eject

{\bf{References}}

\bigskip\bigskip  [1] N.P. Landsman, "Quark Matter 90",
{\it{Nucl. Phys. A}}
 {\bf{525}}, (1991) 397.\smallskip\smallskip\smallskip
 [2] M. Le Bellac, ``Thermal Field Theory" (Cambridge University
Press, 1996). \smallskip\smallskip\smallskip
[3] E. Braaten and R. Pisarski,
{\it{Phys. Rev. Lett.}} {\bf{64}}, (1990) 1338;\smallskip {\it{Nucl. Phys. B}}{\bf{337}}, (1990) 569
.\smallskip J. Frenkel and J.C Taylor, {\it{Nucl. Phys.}} B{\bf{334}}, (1990) 199.\smallskip\smallskip\smallskip
\smallskip 
[4] R.D. Pisarski, {\it{Phys. Rev.
Lett.}} {\bf{63}}, (1989) 1129. \smallskip\smallskip\smallskip 
[5] R. Baier, S. Peign\'e and D. Schiff,
{\it{Z. Phys. C}} {\bf{62}}, (1994) 337;\smallskip P. Aurenche, T.
Becherrawy and E. Petitgirard, hep-ph 9403320 (unpublished).
\smallskip\smallskip\smallskip    
[6] A. Niegawa, {\it{Mod. Phys. Lett.}} A{\bf{10}}, (1995) 379;\smallskip F. Flechsig and A. Rebhan, {\it{Nucl.
Phys.}}
B{\bf{464}}, (1996) 279.\smallskip\smallskip\smallskip
[7] P. Arnold, G.D. Moore and L.G. Yaffee, JHEP {\bf{0206}}, (2002) 030.\smallskip\smallskip\smallskip

[8] B. Candelpergher and T.
Grandou, {\it{Ann. Phys. (NY)}} {\bf{283}}, (2000) 232; \smallskip [arXiv:hep-ph/0009349]; {\it{Nucl. Phys. A}}
{\bf{699}}, (2002) 887.\smallskip T. Grandou, {\it{Acta Physica Polonica B{\bf{32}}}}, (2001) 1185.
\smallskip\smallskip\smallskip

[9] P. Aurenche, F.
Gelis, R. Kobes and E. Petitgirard, {\it{Z. Phys. C}} {\bf{75}}, (1997) 315;\smallskip P. Aurenche, F.
Gelis, R. Kobes and H. Zaraket,\smallskip
{\it{Phys. Rev. D}} {\bf{58}}, (1998) 085003, and references
therein.\smallskip F. Gelis, Th\`ese pr\'esent\'ee \`a l'Universit\'e de Savoie, le 10 D\'ecembre
1998.\smallskip\smallskip\smallskip

[10] T. Grandou, J. Math. Phys. {\bf{44}}, (2003) 611.\smallskip\smallskip\smallskip

[11] T. Grandou, M. Le Bellac and
D. Poizat, {\it{Nucl. Phys. B}} {\bf{358}}, (1991) 408.\smallskip\smallskip\smallskip

[12] T. Grandou, work in progress.
\vfill\eject
{\bf Figure caption}\bigskip\bigskip

{\bf Fig.1:} A graph denoted by $(N,N';1)$, with $N(N')$ insertions of $HTL$ self energy along the
$P(P')$-line, one bare vertex $-ie\gamma_\mu$, and one $HTL$ vertex correction (2.5). It is a $P\!R$ scheme object.
In
the standard $RP$ scheme, the two internal $P(P')$-lines are to be replaced with the full effective propagators
${}^\star S_\alpha(P(P'))$ of Equation (2.2), and this corresponds to the two $RP$ graphs denoted by
$(\star,\star;1)$
in the text. \bigskip

\end